\documentclass[12pt]{article}
\usepackage{amsmath}
\usepackage{graphicx,psfrag,epsf}
\usepackage{enumerate}
\usepackage{natbib}
\usepackage{url} 

\newcommand{\blind}{0}

\begin{document}

\def\spacingset#1{\renewcommand{\baselinestretch}%
{#1}\small\normalsize} \spacingset{1}


\if0\blind
{
  \title{\bf The Midpoint Mixed Model with a Missingness Mechanism (M5):  A Likelihood-Based Framework for Quantification of Mass Spectrometry Proteomics Data (Preprint)}
  \author{Jonathon O'Brien \thanks{
    The authors gratefully acknowledge the Clinical Proteomic Tumor Analysis Consortium (CPTAC) for providing data and guidance in our analysis as well as the National Cancer Institute for supporting this research through the training grant `Biostatistics for Research in Genomics and Cancer', NCI grant 5T32CA106209-07 (T32).}\hspace{.2cm}\\
    Department of Biostatistics, \\University of North Carolina at Chapel Hill\\
    and \\
    Harsha Gunawardena \\
    Department of Biochemistry and Biophysics, \\University of North Carolina at Chapel Hill\\
		and \\
		Xian Chen\\
    Department of Biochemistry and Biophysics, \\University of North Carolina at Chapel Hill\\
		and \\
    Joseph Ibrahim\\
    Department of Biostatistics, \\University of North Carolina at Chapel Hill\\
		and \\
    Bahjat Qaqish \\
    Department of Biostatistics, \\University of North Carolina at Chapel Hill}
  \maketitle
} \fi

\if1\blind
{
  \bigskip
  \bigskip
  \bigskip
  \begin{center}
    {\LARGE\bf The Midpoint Mixed Model with a Missingness Mechanism (M5):  A Likelihood-Based Framework for Quantification of Mass Spectrometry Proteomics Data}
\end{center}
  \medskip
} \fi

\bigskip
\begin{abstract}

Statistical models for proteomics data often estimate protein fold changes between two samples, A and B, as the average peptide intensity from sample A divided by the average peptide intensity from sample B.  Such average intensity ratios fail to take full advantage of the experimental design which eliminates unseen confounding variables by processing peptides from both samples under identical conditions.  Typically this structure is exploited through the estimation of a protein ratio as the median ratio of matched peptide intensities. This simple solution fails to account for a substantial missing data bias which has led to the development of more sophisticated average intensity models. Here we develop the first statistical model that accounts for non-ignorable missingness while utilizing peptide level matched pairs across samples. Our simulation analysis shows that models which fail to utilize peptide level ratios, suffer astonishing losses to accuracy with basic ANOVA estimates having an average MSE 371\% higher than median ratio estimates. In turn, median ratio estimates have an average MSE 35\% higher than our model estimates.  An analysis of breast cancer data reinforces these relationships and shows that our model is capable of increasing the number of proteins estimated by 22\%.
\end{abstract}

\noindent%
\vfill

\newpage
\spacingset{1.45} 
\section{Introduction}
\label{sec:intro}

At the highest level, proteomics is the large scale study of the structure and functions of proteins.  An important class of studies within this field is shotgun, or discovery, proteomics.  These experiments are designed to provide information on a large set of proteins that are not specified before conducting the experiment.  Discovery proteomics experiments typically necessitate the use of a mass spectrometer which entails an inferential step between the readings of the mass spectrometer and the protein level outcomes of interest.  Understanding the details of these experiments becomes essential in order to conduct a sensible analysis of proteomics data.  A statistician might be fooled by the superficial similarities between microarray and proteomics data into simply adopting microarray methods to be used on protein data.  Although the outcomes in the experiments share the same ``intensity'' designation, in reality the similarities end with the name. In a microarray experiment, an intensity refers to the observed brightness of a dye that will be present when a reaction occurs with some target molecule.  The exact relationship between this intensity and the underlying molecular abundance is unclear but, as explained by \cite{Dabney2007}, a researcher at least believes the intensity to be a monotone increasing function of the analyte concentration.  Nothing of the sort can be claimed for intensities from shotgun proteomics experiments.  To justify this statement it will be necessary to provide a basic overview of the shotgun proteomics experiment, including a detailed explanation of the ionization process and the different sources of missing data.  We will show how the experimental details create two statistical features characteristic of all shotgun proteomics experiments; matched pairs data and non ignorable missingness.  After justifying these features we propose a model that accounts for them and test the performance of this model on both simulated and real data.

\subsection{Bottom Up Relative Quantification Experiments} 

Discovery proteomics experiments are usually a type of relative quantification experiment  \citep{Eidhammer2008}.  The  implication  here is that the experiment can only be used to provide  measures  of protein abundance in one sample relative to another, without ever obtaining measures of the absolute abundance in either sample.  Many experiments will compare samples from numerous conditions but the simplest scenario is a comparison of protein abundances between two samples: Sample A and Sample B.  A number of relative quantification workflows are available for proteomics to achieve this goal. These include Label-Free Quantification (LFQ) \citep{Cox2014}, Stable Isotope Labeling by Amino Acids in Cell Culture (SILAC) \citep{Cox2008} and  Isobaric Tags for Relative and Absolute Quantification (iTRAQ) \citep{Ross2004}.  The detailed workflows involved in these experiments are important and should determine the types of model that a statistician would consider.  For instance the model proposed in this paper works for SILAC and LFQ data, but the missing data mechanism we employ is inappropriate for an iTRAQ experiment.  Nonetheless a full explanation of each experimental workflow is not necessary for our purposes.  The experiments described here are referred to as bottom-up proteomic methods, because proteins are too large to be identified by mass which forces us to make inference about relative protein abundance from measurements obtained on amino acid fragments called peptides. A typical bottom-up proteomic workflow involves the extraction of proteins from cells, tissues or biological secretions/fluids, followed by proteolysis which breaks proteins into peptides.  Typically this is done by adding an enzyme called trypsin that specifically cleaves proteins at lysine and arginine amino acid residues.  After this digestion occurs peptides from the sample are separated according to each peptide's hydrophobicity in a process called elution, where the more hydrophobic peptides will be the last to separate. After elution, peptides will travel towards an ionization device which converts the peptides into ions so that they may enter and be manipulated by a mass spectrometer.  The ionization is usually done with electricity in the form of electrospray ionization (ESI) or with lasers by matrix-assisted laser desorption/ionization (MALDI).  It is important to emphasize that both of these ionization technologies affect large numbers of peptides all at the same time and not all of them will successfully ionize.  The elution process aims to completely separate each group of peptides but it does not work perfectly.  When referring to the measurement made on a specific peptide we may refer to all of the other pepties that were ionized at the same time as co-eluting peptides.  Co-eluting peptides play an important role in determining the probability of ionization.  After ionization the newly formed peptide ions are manipulated by a mass spectrometer which is capable of separating the ions according to their mass and counting the number of ions corresponding to each mass.  Peptides with a large counts will be selected for fragmentation and a second mass spectrometry reading of the fragments.  The process of selecting peptides for a second mass spectrometry step based on the relative magnitude of the counts is called data-dependent analysis.  The second mass spectrometry step is mostly used to identify the peptides that were just counted (quantification also happens during this step in an iTRAQ experiment). A summary of the ion counts for each now identified peptide, usually computed as the area under a curve from a plot of counts through time \citep{Cox2008}, is referred to as a peptide intensity.  A more comprehensive description of the LFQ workflow can be found in a paper by \citet{Sandin2011}.  For the purposes of this paper we will focus on only the experimental details which motivated our statistical model.

\subsection{Matched Pairs Data}
Mass spectrometers work with ions because advances in technology have given us a tremendous ability to manipulate ions.  For this reason ionization of peptide molecules is an indispensable aspect of a mass spectrometry proteomics experiment.  Unfortunately, and this is a critical point, not all of the peptides from the sample will be ionized. Certain peptides tend to ionize more efficiently, while others will not ionize at all.  The probability that a given peptide molecule will be ionized can be referred to as ionization efficiency. Ionization efficiency is believed to be a property of both the chemical structure of each peptide and the presence of other co-eluting peptides, sometimes referred to as matrix interferences.   \citet{Schliekelman2014} found that competition for charge between background peptides may actually be a more important factor than abundance in determining if a peptide will be detected.  Regardless of what factors are most important, ionization efficiency can cause the proportion of peptides that enter into the mass spectrometer to be drastically altered.  This is why we previously claimed that peptide intensities are not a monotone increasing function of peptide concentrations.  One peptide might be far more abundant than another in the original sample but a lower ionization efficiency could reverse the relationship for peptide intensities.  What we observe is the number of peptides in solution multiplied by that peptide's ionization efficiency.  Fortunately, if the efficiency parameter is a property of the individual peptide, it will cancel out when put into a ratio with the same peptide from the other sample.  This relationship is outlined in Table~\ref{tab:tabone}.
\begin{table}
\caption{This table shows the relationship between relative protein abundance and the intensities of a peptide belonging to that protein. $p$ is the probability that the peptide ionizes and makes its way into the mass spectrometer.  $pW$ and $pZ$ represent the expected intensities from samples A and B respectively. 
\label{tab:tabone}}
\begin{center}
\begin{tabular}{rrrr}
 & Protein Abundance & Peptide Abundance & Ion Abundance \\\hline
Sample A & $X$ & $W$ & $pW$  \\
Sample B & $Y$ & $Z$ & $pZ$ \\
Ratio & $\frac{X}{Y} = \mu$ & $\frac{W}{Z} = \frac{X}{Y} = \mu $& $\frac{pW}{pZ}=\mu $\\
\end{tabular}
\end{center}
\end{table}
In a SILAC experiment every peptide in both samples will be processed at the same time and thus will be exposed to the same conditions yielding identical ionization efficiencies.  However, even in a SILAC replicate the efficiency may be altered due to variations in sample preparation and elution time resulting in a different profile for the background peptides.  For a Label-Free experiment, run to run variation should always be expected.  So unless the researchers have good reason to assume the ionization efficiencies will be equivalent, the outcomes may be confounded by run to run variation affecting each peptide differently. The incomplete and inconsistent ionization of analytes makes it impossible to accurately measure the abundance of proteins from the original sample. However in ratio form we can still make inference about the relative abundance of proteins in two samples. This is why proteomics experiments are often referred to as relative quantification experiments and it is why popular proteomics software packages, such as MaxQuant \citep{Cox2008} estimate log protein ratios as the median of log peptide ratios.  Similar methods dominate the techniques discussed by \cite{Eidhammer2008}.  Despite the ubiquity of median ratio estimates in proteomics software packages, none of the models we found in the literature make use of peptide level ratios beyond including peptide as a covariate in a linear model.  Instead, most statistical methods estimate the log protein ratio from Sample A to B by taking the average log intensity across all peptides within the protein from Sample A and then subtracting the average log intensity from all peptides within the protein from sample B.  In the absence of missing data these methods are almost identical since $E(X-Y)=E(X)-E(Y)$. However, in the presence of wide-spread missing data many peptides are detected in only one of the two samples.  This makes it difficult to interpret the results from a method based on average intensities.  Unfortunately wide-spread missingness is unavoidable in a discovery mass spectrometry experiment as explained below. 

\subsection{Intensity-Dependent Missingness}
Unlike microarray experiments in which missing values often comprise about 1-11\% of the data \citep{DeBrevern2004}, proteomics data sets almost always have a much higher percentage of missing data.  In the dataset analyzed in this paper 25\% of the peptides were missing and according to \citet{Karpievitch2009}, 50\% missing values are not uncommon.  For this reason, the way we conceptualize and treat the missing data will take on huge importance.  Here we will briefly discuss some of the largest causes of missing data.
\subsubsection{Detection Limit}  Mass spectrometers have both theoretical and a practical limits of detection (LOD).  The  theoretical LOD is the minimum number of ions a given instrument can capture and produce an ion current with adequate signal enhancement.  Although any peptide exceeding this number of ions can theoretically be detected by the mass spectrometer every sample contains far more than one peptide which results in a considerable amount of noise.  This noise results in a practical detection limit, which is dependent on the sample itself, whereby the software fails to distinguish peptide peaks from background  noise.  For this reason, sample-related factors that either result in a higher practical detection limit or a decreased intensity due to the nature of the sample can result in missing values. As previously discussed, a major driver in this setting is the peptide ionization efficiency.  If the ionization efficiency is low then the intensity will be low and may fall below the detection limit.  This is clearly a form of non-ignorable missingness where the probability of being missing is directly related to the magnitude of the intensity.
\subsubsection	{Data-Dependent Tandem Mass Spectrometry}  
Most mass spectrometers performing data-dependent analysis (DDA) do not succeed in fragmenting every ionized peptide. Peptides are mass selected (isolated) for a tandem MS (MS/MS) step according to their intensity rank order.  This tandem mass spectrometry step is where peptide identification occurs, so any peptide signal that is not selected for tandem mass spectrometry will not yield useful data.  On a side note, in an iTRAQ experiment quantification also occurs in the tandem step whereas in LFQ and SILAC experiments quantification occurs during the first MS step. This is why our proposed missing data mechanism does not apply to iTRAQ experiments.  The data that we analyze in this paper was generated from a Q Exactive mass spectrometer made by Thermo Scientific, which was only capable of capturing approximately 80\% of the peak intensities above the LOD. Thus even above the practical LOD an intensity dependent process can result in missing values.  The number of detectable (identifiable) features can be increased with advances in the operating speed of the device, and complete parallelization can be achieved in newer Orbitrap instruments \citep{Lesur2015}. However, this results in a trade off where the identification process becomes less certain. In most settings we will have to consider two sources of non ignorable missingness, one which occurs below a frequently changing detection limit and another above.

\subsubsection{Misidentified/Unmatchable/Razor Peptides}  A peptide might appear in one sample and not in another simply because it was misidentified.  Matching algorithms are designed to minimize this problem but it undoubtedly still exists.  A similar problem comes from razor peptides.  These are peptides that are properly identified but that could belong to more than one protein.  Many software programs assign razor peptides to the protein that has the most peptide level evidence based on the Occam's Razor concept of protein parsimony \citep{Cox2008}.  Yet this process could result in misidentification and consequently, missing values.  It is also possible that a particular peptide will simply fail to be identified with any certainty which will also result in missing values.  It is probably safe to classify missingness caused by classification errors as missing at random.  

\subsubsection{Modeling Missingness} Many efforts have been made to correct for missing data biases in mass spectrometry experiments.  A review of missing data techniques in proteomics by \cite{Taylor2013} compared three methods for removing missing data bias: an accelerated failure time (AFT) model by \cite{Tekwe2012}, a mixture model proposed by \cite{Karpievitch2009} and K-Nearest Neighbors (KNN) imputation as described by \cite{Troyanskaya2001}.  Notable additions to this group include three Bayesian methods. One method proposed by \cite{Luo2009} models missing probabilities with a logit function. The Bayesian model of \cite{Lucas2012} attempts to account for missingness caused by misidentification and \cite{Koopmans2014} proposes a model which allows for a random detection limit.  However, none of these methods utilize peptide level ratios in their solutions which leaves room for improvement.  On their merits as purely missing data techniques, only the model by \cite{Luo2009} theoretically accounts for all the sources of missingness described above.  The AFT model and the mixture model both assume a fixed detection limit.  But we should expect this detection limit to vary from peptide to peptide depending on what other compounds are being processed in the background.  Furthermore, both of them assume that any missingness above this theoretical detection limit should be missing at random with the mixture model explicitly categorizing all missingness as either at random or due to a detection limit.  The use of any technology which uses Data Dependent Analysis will make this assumption false since data dependent analysis is a form of intensity dependent missingness that occurs above the detection limit.  The interesting effort by \cite{Koopmans2014} which attempts to model a random detection limit analyzes data at the protein level.  In other words, the estimation within an experiment has already occurred and it is unclear what missing data bias can be corrected. As for the KNN approach, it is difficult to imagine that there even could be a theoretical justification for imputing 40\% of a dataset and then proceeding with an analysis as though the observations were real.  Nonetheless, this option has made its way into software packages such as Inferno (Formerly called Dante) described by \cite{Polpitiya2008}, so we will examine its efficacy later.  The problem of ascertaining the cause of a missing value is probably intractable.  However, we can say with some confidence that the probability of an intensity being missing should be a monotone increasing function of the intensity.  Although the exact conditions and sources of missingness will vary from peptide to peptide and experiment to experiment a single monotone parametrized function of the probability of missingness could serve as a useful approximation for the conglomeration of missing data sources.  For this reason, along with some mathematical niceties, we model a probit missing data mechanism such that for each peptide the probability of being observed is given by $\Phi(a+by)$ where $\Phi$ is the standard normal CDF, $y$ is the peptide intensity, and $a$ and $b$ are missingness parameters to be estimated in our analysis.  

\section{Methods}
\label{sec:meth}
In order to symmetrically model log peptide intensities, for the rest of the paper we will refer to intensities only on the log scale, for each peptide we frame the problem in terms of the protein fold change (the difference between the intensities) and the midpoint of the two peptides, 
$$
  Y_{ijk} \sim N(\alpha_{j(i)} +(-1)^{k+1}\frac{\mu_{i}}{2},\sigma) \\
$$

where $Y_{i,j,k}$ is the intensity of the $j$th peptide within the $i$th protein from the $k$th sample, $k = 1,2$, $i =1,\dots,n$ indexes the unique proteins in the samples and $j =1,\dots,m_{i}$ indexes the peptides within the $i$th protein, $\alpha_{j(i)}$ represents the midpoint of the two intensities of peptide $j$ within protein $i$ and $\mu_{i}$ represents the protein fold change.  The notation $N(\beta,\sigma)$ denotes a normal random variable with mean $\beta$ and variance $\sigma$.  The mixed model definition is completed with $\alpha_{j(i)} \sim N(\beta_{\alpha},\xi)$ independent from $\mu_{i} \sim N(\beta_{\mu},\tau)$.

This midpoint mixed model (M3) provides a symmetric framework for analyzing a proteomics experiment in one statistical model while accounting for the matched pairs nature of the data.  The model can easily be fit using standard software such as PROC MIXED in SAS or lme from the NLME package in R.  We expect this model to provide similar results to standard ratio-based methods and improved downstream analysis by creating a single estimate of experimental variance.  In fact, if we used fixed effects in place of random effects M3 is a reparameterization of a linear model with covariates for peptide within protein, sample, and a sample*protein interaction.  Of course this model completely fails to account for missing data bias.  We expand the M3 model into a selection model with a probit missingness mechanism. We refer to this new model as the midpoint mixed model with a missingness mechanism (M5).  Let $I()$ be an indicator function so that $R_{ijk}=I(Y_{ijk} \mbox{ is observed})$. We assume $(R_{ijk} | Y_{ijk} )\sim Bernoulli(\Phi(a+bY_{ijk}))$.   

Fitting this model is greatly complicated by the number of missing values in a proteomics experiment.  In our dataset, there were certain proteins with over 200 missing values, and integrating the likelihood 200 times created computational difficulties.  We resolved this issue by giving each parameter a non-informative prior and using a Gibbs Sampler.  The Gibbs Sampler required three sampling steps that were non-standard: the distribution of a missing value given everything else $f_{(Y_{ijk}|\mu_{i},\alpha_{i},Y_{ijk'},\mathbf{\theta},\mathbf{R})}$, where $Y_{ijk'}$ is the matched pair corresponding to the missing value and $\mathbf{\theta}$ is the vector of parameters $(a,b,\tau,\xi,\sigma,\beta_{\alpha},\beta_{\mu})$; the distribution of a protein fold change given everything else $f_{(\mu_{i}|\alpha_{i},\mathbf{Y},\mathbf{\theta},\mathbf{R})}$; and the distribution of a midpoint given everything else $f_{(\alpha_{j(i)}|\mu_{i},\mathbf{Y},\mathbf{\theta},\mathbf{R})}$.
A bit of manipulation reveals that the distribution of a missing value follows the Extended Skew Normal distribution as described by \cite{Azzalini2014}.
$$
f_{(Y_{ijk}|\mu_{i},\alpha_{i,j},Y_{ijk'},\mathbf{\theta},\mathbf{R})}(x)=\frac{\phi(\frac{x-\mu_{x}}{\sqrt{\sigma}})\Phi(-a-bx)}{\sqrt{\sigma}\Phi(\omega)}
$$
where
$$
\mu_{x}=\alpha_{j(i)} + (-1)^{k+1} \frac{\mu_{i}}{2},\quad \omega=\frac{-a-b\mu_{x}}{\sqrt{1+\sigma b^2}}.
$$

We also find that 
$$
(\mu_{i}|\alpha_{i},\mathbf{Y},\mathbf{\theta},\mathbf{R})\sim N\left(\frac{\beta_{\mu}\sigma+\frac{\tau}{2}\sum_{j}{(y_{ij1}-y_{ij2})}}{\sigma+\frac{m_{i}\tau}{2}},\frac{\sigma\tau}{\sigma+\frac{m_{i}\tau}{2}}\right),
$$
and
$$
(\alpha_{ij}|\mu_{i},\mathbf{Y},\mathbf{\theta},\mathbf{R})\sim N\left(\frac{\beta_{\alpha}\sigma+\xi(y_{ij1}+y_{ij2})}{\sigma+2\xi},\frac{\xi\sigma}{\sigma+2\xi} \right).
$$
Proofs of these results can be found in Appendix~\ref{sec:proofs}.
Although these formulas are complex the results are somewhat intuitive.  Each missing value from a pair of points comes from a skew normal distribution where the skew is determined by the missing data mechanism.  The fold change comes from a distribution centered around a weighted average of the mean protein fold change and the average of the observed differences in peptide intensities.  The peptide midpoint is drawn from a distribution centered around a weighted average of the mean peptide midpoint and the observed midpoint of the pair of intensities.  

The Bayesian model formulation is completed with the priors in Table~\ref{tab:priors}.
\begin{table}
\caption{The prior and posterior distributions used to complete the model.  The parameters with non standard prior and posterior distributions are described in the text.}
\label{tab:priors}
\begin{tabular}{|l|c|c|}
  \hline
  Parameter & Prior & Posterior \\
  \hline \hline
 $\tau$ & $InverseGamma(.001,.001)$ & $InverseGamma(.001+n/2,.001+\frac{\sum{\mu_{i}-\beta_{\mu}}}{2})$ \\
 $\xi$ & $InverseGamma(.001,.001)$ & $InverseGamma(.001+\sum{m_{i}}/2,.001+\frac{\sum{\alpha_{i}-\beta_{\alpha}}}{2})$ \\
 $\sigma$ & $InverseGamma(.001,.001)$ & $InverseGamma(.001+\sum{2*m_{i}}/2,.001+\frac{\sum{\epsilon_{i}}}{2})$ \\
 $\beta_{\alpha}$ & $N(0,10000)$ & $N(\sum{\alpha_{i,j}}/\xi,(\frac{1}{10000}+\frac{\sum{m_{i}}}{\xi}))$ \\
 $\beta_{\mu}$ & $N(0,10000)$ & $N(\sum{\mu_{i}}/\tau,(\frac{1}{10000}+\frac{\sum{n}}{\tau}))$ \\
 $a$ & $N(0,10000)$ & Probit Regression Estimation \\
 $b$ & $N(0,10000)$ & Probit Regression Estimation \\
  \hline
\end{tabular}
\end{table}
The posterior distribution of $(a,b)$ is estimated by fitting the probit regression model 
$$
\Phi^{-1}(E[R_{ijk}|y_{ijk}])=a+by_{ijk}
$$ 

The posterior distribution is then approximated as 
$$
\begin{pmatrix} a\\ b \end{pmatrix} \sim N(\begin{pmatrix} \hat{a}\\ \hat{b} \end{pmatrix},\hat{\Sigma})
$$
Where $\hat{a}, \hat{b}$ and $\hat{\Sigma}$ are the parameter estimates from the probit regression and their corresponding covariance estimate, respectively.  The bivariate normal distribution used here approximates the posterior distribution as a consequence of Bayesian large sample theory \citep[chap. 4]{Gelman2004}. 

\subsection{Simulation Study}
To explore the potential benefits of the M5 model, we conduct simulations to compare the accuracy of our estimates to six other estimation procedures.  In addition to the M3 and M5 models, we analyze the commonly used method of median ratios, a one-way ANOVA for protein within sample, QRollup and QRollup performed after implementing a weighted K-Nearest Neighbors imputation (KNNQ).  QRollup is a method of analyzing average intensities that seeks to avoid missing data bias by analyzing only the largest 66\% of the intensities within each protein/sample combination.  A software package called Inferno implements the QRollup method\citep{Polpitiya2008}. Similar approaches are discussed by \cite{Eidhammer2008}.  The Inferno software also offers an option to use Weighted K-Nearest Neighbors Imputation, which is why we coupled imputation with the QRollup method.  The ANOVA model is intentionally simpler than that described by \cite{Oberg2012}, as we wanted to use an example of a model where proteins are estimated as the average intensity within groups regardless of the presence of a matched pair.  Linear models that include peptide as a covariate should perform similarly to the M3 model.  This set of methods gives us a look at the performance of three methods based on ratios and three methods based on average intensities.  All methods, except for M3 and M5, are currently supported by proteomic software packages.   

In our simulation the hyperparameter values were 
$$
\tau=9, \xi=4, \sigma=.3, a=-9, b=.5, \beta_{\alpha}=18.5, \beta_{\mu}=0.
$$
We generated 500 protein fold changes from a $N(\beta_{\mu},\tau)$ distribution and generated the number of constituent peptides within each protein by sampling with replacement from the set $\{1,...,12\}$.  For each peptide we generated independent random midpoints from a $N(\beta_{\alpha},\xi)$ distribution.  We generated independent residual errors, $\epsilon_{ijk}$ as $N(0,\sigma)$ random variables.  Then we created intensities $y_{ijk}=\alpha_{j(i)} +(-1)^{k+1} \frac{\mu_{i}}{2}+\epsilon_{ijk}$.  Next we simulated missingness by computing probabilities of missingness for each intensity as $p_{ijk}=\Phi(a+by_{ijk})$, then we randomly drew Bernoulli random variables, $(R_{ijk})$, according to those probabilities, to identify which $y_{ijk}$ are missing.  We then fit all six models including 1,000 draws from the M5 Gibbs Sampler to create M5 estimates.  Results were recorded and the whole process was repeated 100 times.

\subsection{Protein Categories}
\begin{figure}
\begin{center}
\includegraphics[width=0.8\textwidth,natwidth=750,natheight=456]{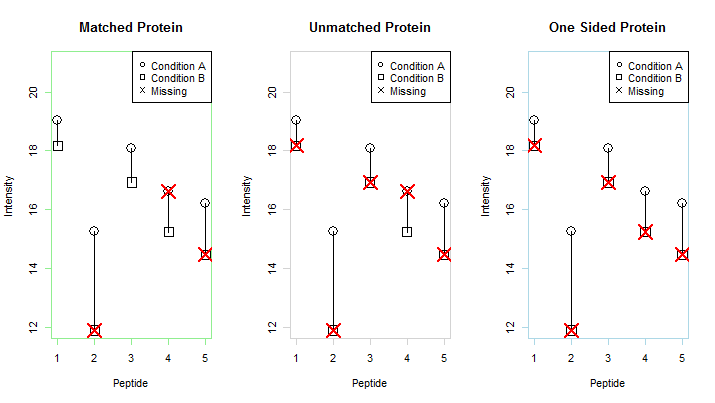}
\end{center}
\caption{The three categories of proteins.  Matched proteins contain at least one matched peptide pair.  Unmatched proteins contain data from both conditions but no matched pairs. One-sided proteins contain peptide measurements from only one sample. \label{fig:ProtCategory}}
\end{figure}

Before comparing the six methods, some classification of observation patterns is needed since not all methods are capable of estimating the same proteins.  To this end we classify each protein as ``matched'', ``unmatched'', ``one-sided'' or ``missing''. Figure~\ref{fig:ProtCategory} presents a visual depiction.  

Missing proteins are proteins for which peptides were identified but no peptide intensities were observed.  These are not interesting and even though they can be estimated with M3, M5 or KNNQ we recommend just removing them from the study.  Matched proteins are proteins that have at least one matched peptide pair.  With at least one shared peptide from each sample, all of the methods can be used for estimation.  An unmatched protein has observed intensities from each sample but no peptides that are quantified in both samples.  One-Sided proteins have intensities from peptides in only one sample and are completely missing in the other. This can be indicative of a large fold change difference.  M5, M3 and KNNQ can be used to estimate all types of proteins.  The ANOVA model and QRollup can be used for both matched and unmatched proteins while the method of median ratios can only be used on matched proteins.

\subsection{Simulation Results}
The sampling chains all appeared to achieve stationary distributions after about 300 draws (in the real data this was achieved within 50).  For this reason, our estimates were based on the posterior mean after a burn in length of 500 draws. 

\begin{figure}
\begin{center}
\includegraphics[width=0.8\textwidth,natwidth=717,natheight=436]{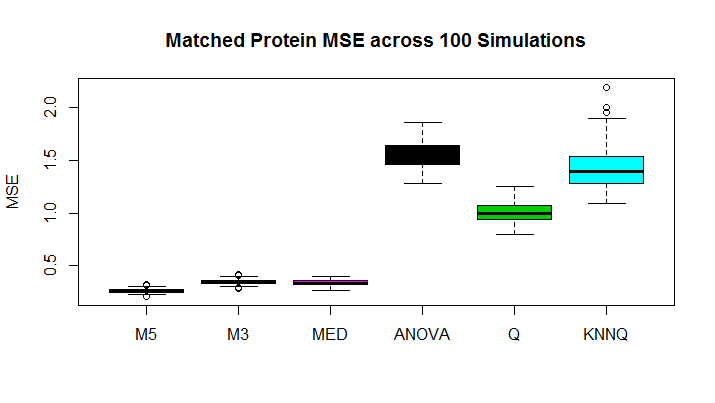}
\end{center}
\caption{MSE for each method computed across matched Proteins and within each simulation. \label{fig:matchedbox}}
\end{figure}
\begin{figure}
\begin{center}
\includegraphics[width=0.8\textwidth,natwidth=900,natheight=546]{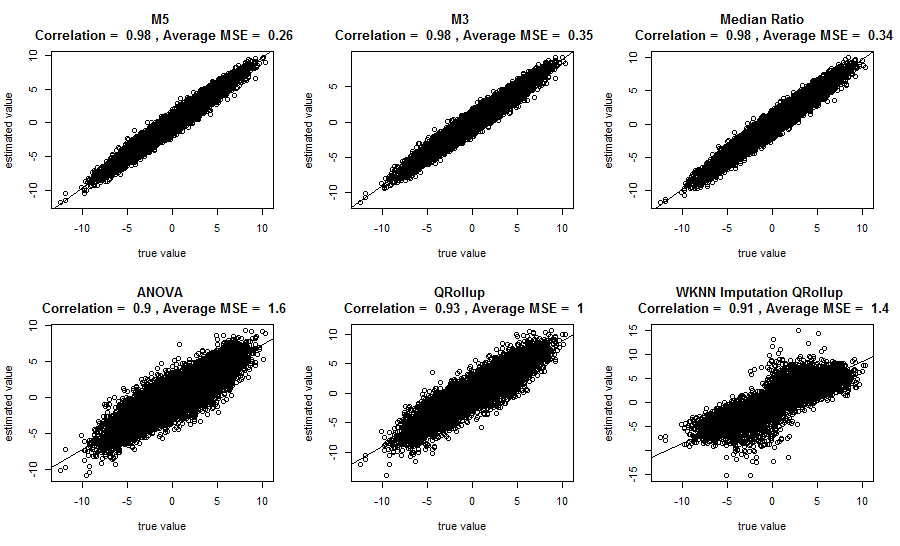}
\end{center}
\caption{Scatterplot of true simulated fold changes for matched Proteins vs their estimates across all simulations.  Correlation coefficients are also computed across all simulations. \label{fig:matchedscatter}}
\end{figure}

Figure~\ref{fig:matchedbox} shows the distribution of mean squared errors across simulations.  The most obvious result here is that the methods based on ratios are far outperforming methods based on average intensities.  M5 demonstrates the best performance with an average MSE of 0.26.  The method of medians was the second most accurate with an MSE of 0.35, which represents an increase in error of 35\%.  This is a fairly large increase but it is hardly noticeable relative to the error coming from the average intensity methods.  The best of these was QRollup with an average MSE of 1, which represents an increase of 285\%.  
It should be noted that the commonly used validation tool of correlation does not do a very good job of assessing algorithmic weaknesses here.  Figure~\ref{fig:matchedscatter} shows that even though some of these methods more than sextuple mean squared error, the lowest correlation coefficient is still above 0.9. It should also be noted that the use of Weighted K-Nearest Neighbors appears to be detrimental to the accuracy of QRollup estimates.  

\begin{figure}
\begin{center}
\includegraphics[width=0.8\textwidth,natwidth=717,natheight=436]{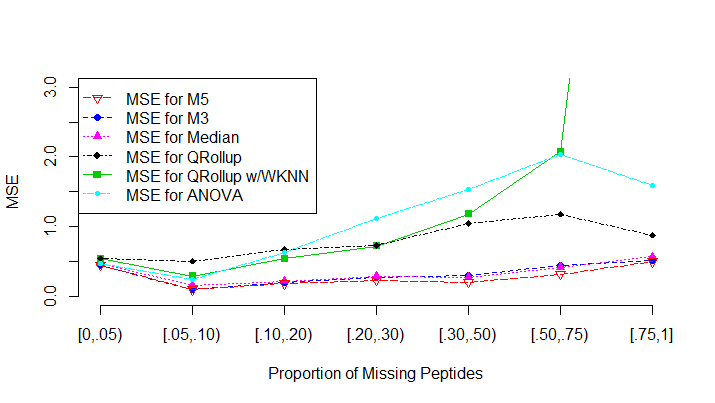}
\end{center}
\caption{MSE for each method according to categories of the percentage of missing peptides.  The MSE for KNNQ at 75\% missing data is 12.81 which was too extreme to be plotted along with the other methods. \label{fig:bymiss}}
\end{figure}

These relationships can be further explored by categorizing proteins according to the percentage of peptides which are missing as shown in Figure~\ref{fig:bymiss}. This plot shows that the error for the KNNQ method increases substantially once more than 50\% of the data requires imputation.  In this chart we can see that, as missingness increases, the ANOVA estimation also loses accuracy at a much faster rate than the other methods.  This is likely because  the ANOVA model simply reports average intensities within each sample regardless of the amount of missing data.  

%

\begin{figure}
\begin{center}
\includegraphics[width=0.8\textwidth,natwidth=717,natheight=436]{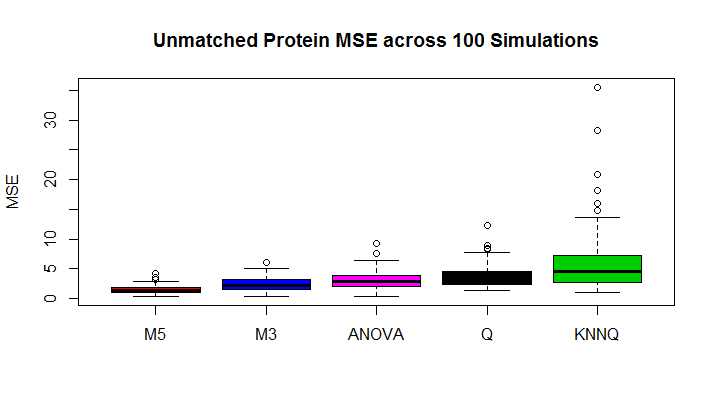}
\end{center}
\caption{MSE for each method computed across unmatched proteins and within each simulation.  The method of medians, MED, is not applicable to unmatched proteins.  \label{fig:unmatchedbox}}
\end{figure}

In the case of one-sided and unmatched Proteins the method of medians is obviously not applicable. Among the other methods the rank ordering based on average MSE remains the same. (see Figure~\ref{fig:unmatchedbox}).  

\begin{figure}
\begin{center}
\includegraphics[width=0.8\textwidth,natwidth=900,natheight=547]{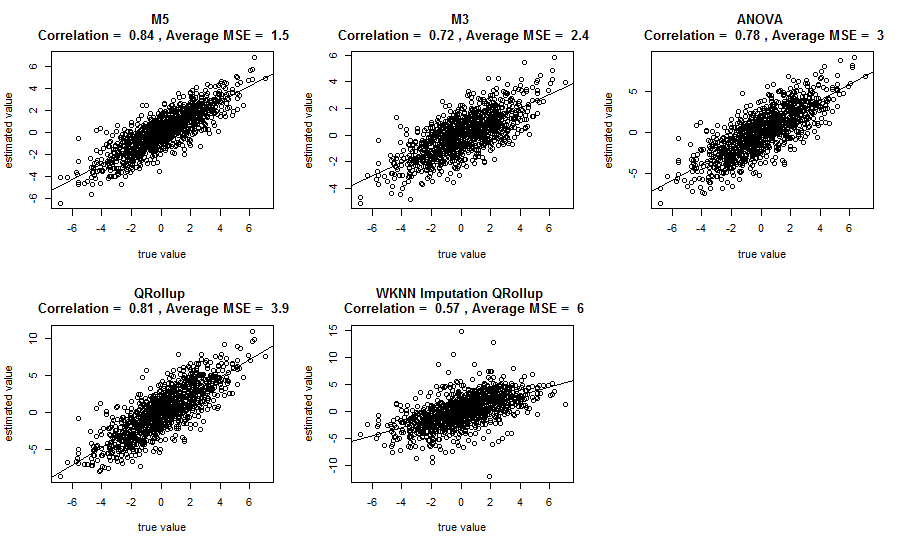}
\end{center}
\caption{Scatterplot of true simulated fold changes for unmatched Proteins vs their estimates across all simulations.  Correlation coefficients are computed across all simulations. \label{fig:unmatchedscatter}}
\end{figure}

In this case the average MSE for M5 is 1.5 and the second best is the M3 model at 2.4.  The best average intensity method was the ANOVA model with an MSE of 3.  Correlation coefficients are much weaker in this category as pictured in Figure~\ref{fig:unmatchedscatter}.

\begin{figure}
\begin{center}
\includegraphics[width=0.8\textwidth,natwidth=717,natheight=436]{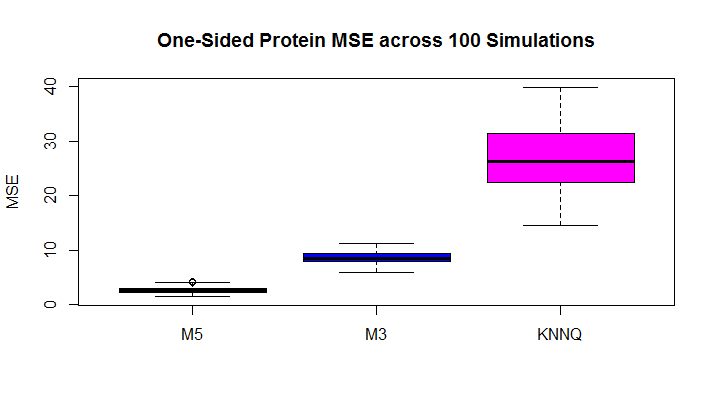}
\end{center}
\caption{MSE for each method computed across One-Sided proteins and within each simulation.  The method of medians (MED), ANOVA, and QRollup (Q) are not applicable to One-Sided proteins  \label{fig:onesidedbox}}
\end{figure}

\begin{figure}
\begin{center}
\includegraphics[width=0.8\textwidth,natwidth=900,natheight=547]{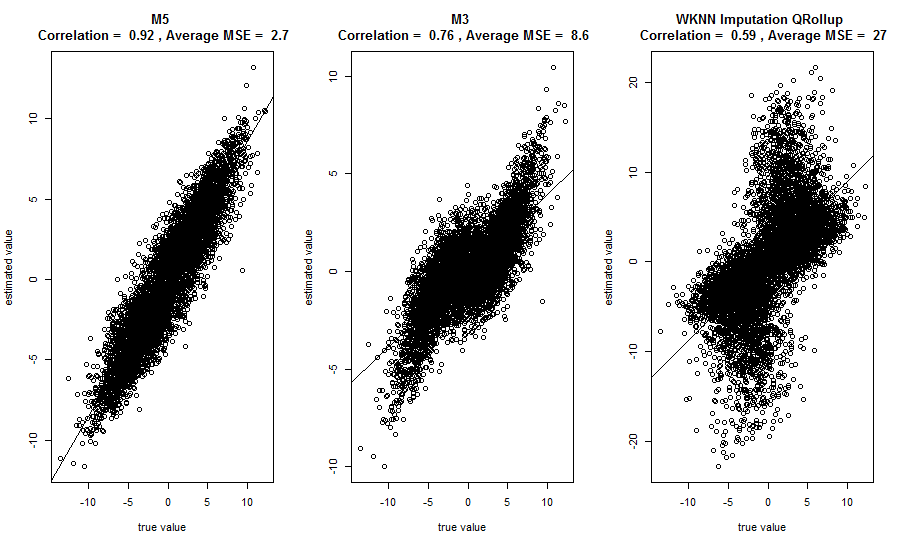}
\end{center}
\caption{Scatter plot of true simulated fold changes for one-sided proteins vs their estimates across all simulations.  Correlation coefficients are computed across all simulations. \label{fig:onesidedscatter}}
\end{figure}

Arguably the greatest advantage to using the M5 model comes from the ability to estimate One-Sided proteins.  These proteins are difficult to estimate since one of the samples provides no observed values.  Keep in mind for a One-Sided protein that we could assume that the abundance of the missing value is between zero and the detection limit.  However, the upper bound on an abundance ratio is infinite.  Nonetheless, M5 does provide decent estimates in this situation as can be seen in Figure~\ref{fig:onesidedbox}.  
Only three methods were capable of estimating one-sided proteins and only one of them could be considered useful.  The MSEs for M5, M3 and KNNQ were respectively 2.7, 8.6 and 27.  The range of the log-scale fold changes in this simulation were roughly -10 to 10.  So an average MSE for M5 of 2.7 is certainly small enough for the estimates to be of interest.  The scatter plot in Figure~\ref{fig:onesidedscatter} strongly highlights the advantages of the M5 model.

\section{Breast Cancer Data}
\label{Chapter:RealData}
\begin{figure}
\begin{center}
\includegraphics[width=0.8\textwidth,natwidth=717,natheight=436,trim = 0mm 50mm 0mm 0mm]{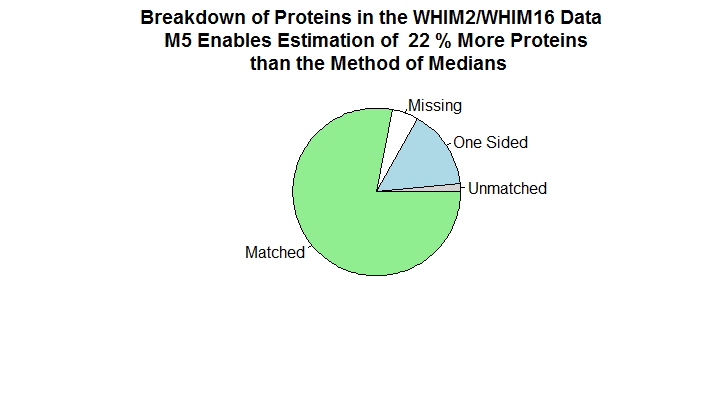}
\end{center}
\caption{Distribution of proteins in the tumor data. \label{fig:piechart}}
\end{figure}

In order to make sure the results of our simulation study are not artifacts of the data generation procedure, we also analyzed the effect of non-informative missingness on a real data set.  
The data, generated by the Chen Biochemistry Lab, contains peptide level LFQ measurements from two samples of breast cancer tissue (one Basal and one Luminal).  This dataset can be found in the supplementary material.
11,866 unique proteins were identified in the data, of these 594 were Missing, 9,265 had at least one peptide pair, 1,810 were one-sided and 197 had intensities in both samples but no matched pairs.  This breakdown is pictured in Figure~\ref{fig:piechart}.  

\begin{figure}
\begin{center}
\includegraphics[width=0.8\textwidth,natwidth=715,natheight=403]{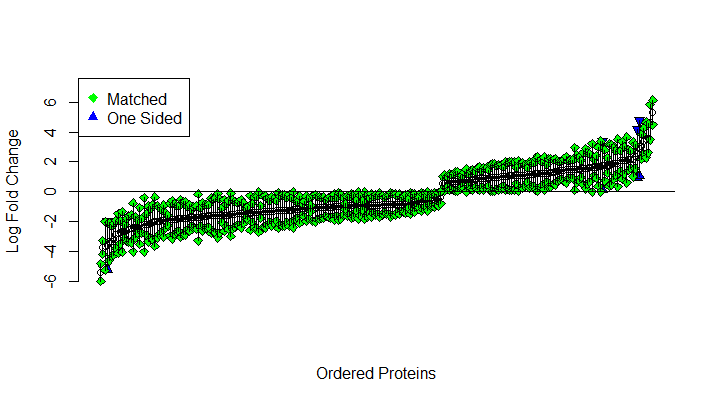}
\end{center}
\caption{M5 estimates of the log fold change between proteins found in Basal and Luminal breast cancer tissues.  Only proteins with 95\% credible intervals that do not contain zero are pictured. \label{fig:catepillar}}
\end{figure}

Not considering missing proteins, we can see that before we even do an analysis the M5 model is capable of estimating an additional 2,007 (22\%) proteins compared with the method of medians.  This would be a substantial gain if our method is capable of estimating those proteins with a decent level of accuracy as our simulations suggest.  Furthermore, the entire data set contains information on 248,342 peptides, 61,418 (about 25\%) of which are missing.  There is a tremendous amount of information in the patterns of those 61,418 missing values, and in theory the M5 model takes full advantage of them.  
M5 model estimates were computed on a random set of 1,000 proteins and 95\% credible regions were computed.  1,000 draws from the gibbs sampler were used with a burn in length of 500.  We reduced the data size purely for computational simplicity.  From the 1,000 proteins, 252 matched Proteins and 4 one-sided Proteins did not contain zero in their credible intervals (Figure~\ref{fig:catepillar}).  Among the 4 one-sided Proteins is protein O75363 which is better known as the gene product for the Novel Amplified in Breast Cancer-1 gene (NABC1).  This gene is known to be involved in cancer typically being upregulated in breast cancers and downregulated in colon cancer \citep{Beardsley2003}.  Since this protein was one-sided in our dataset no useful information regarding NABC1 would have been found without the M5 model.
After estimation, we computed the square of the M5 estimates divided by the posterior standard deviation.  The proteins were then ranked in descending order and are presented in Table~\ref{tab:results}.
\begin{table}
\caption{Twenty proteins ordered by the highest squared ratio of posterior mean to posterior standard error.  At the bottom of the table is the complete set of one-sided proteins for which the credible region did not include zero.}
\label{tab:results}
\begin{tabular}{|l|c|c|c|}
  \hline
  Protein & Estimate & Posterior SD & Category \\
  \hline \hline
        P48681 & -5.39 & 0.300 & Matched \\
        O95425 & -3.72 & 0.230 & Matched \\
        O76070 & 5.31 & 0.420 & Matched \\
        Q13557 & -2.64 & 0.231 & Matched \\
        F8VTL3 & -1.89 & 0.169 & Matched \\
        Q07065 & -2.33 & 0.211 & Matched \\
        Q13813 & -0.97 & 0.096 & Matched \\
        P12270 & -1.12 & 0.112 & Matched \\
        G3XAI2 & -2.26 & 0.241 & Matched \\
        Q9Y4L1 & -1.42 & 0.163 & Matched \\
        P97457 & 4.653 & 0.597 & Matched \\
        Q86SF2 & 3.39 & 0.436 & Matched \\
        O60231 & -1.93 & 0.25 & Matched \\
        Q13363 & 3.58 & 0.501 & Matched \\
        Q9NX62 & 2.16 & 0.302 & Matched \\
        Q5T6V5 & 3.02 & 0.426 & Matched \\
        Q86WJ1 & -2.70 & 0.386 & Matched \\
        P23786 & 1.77 & 0.261 & Matched \\
        Q6BCY4 & -2.55 & 0.376 & Matched \\
        Q9NP74 & -2.37 & 0.357 & Matched \\ 
				\hline
        P12109 & -3.59 & 0.813 & One-sided \\
        Q16666 & 1.76 & 0.823 & One-sided \\
        B4E1Z4 & 2.55 & 0.817 & One-sided \\
        O75363 & 2.90 & 0.944 & One-sided \\
  \hline
\end{tabular}
\end{table}
We also used the real data to study the effects of non-ignorable missingness on each of the six estimation methods.  To accomplish this goal we first reduce the data to allow a complete case analysis, so that only peptides with observed intensities from both samples are included in the reduced dataset.  From this complete-case data, 500 proteins were randomly selected for a sensitivity analysis. The mean peptide ratio within each protein was calculated and considered to be the reference value.  We explored what happens to the estimates from each model as higher levels of intensity-dependent missingness are introduced.  Appropriate values of the missingness parameter $b$ were discovered by trial and error to provide overall missingness levels of 1, 5, 10, 20, 30, 40 and 50 percent.  Mean squared error was then calculated for each of the six methods on all 7 datasets.  

\subsection{Results of the Sensitivity Analysis}
\begin{figure}
\begin{center}
\includegraphics[width=0.8\textwidth,natwidth=717,natheight=436]{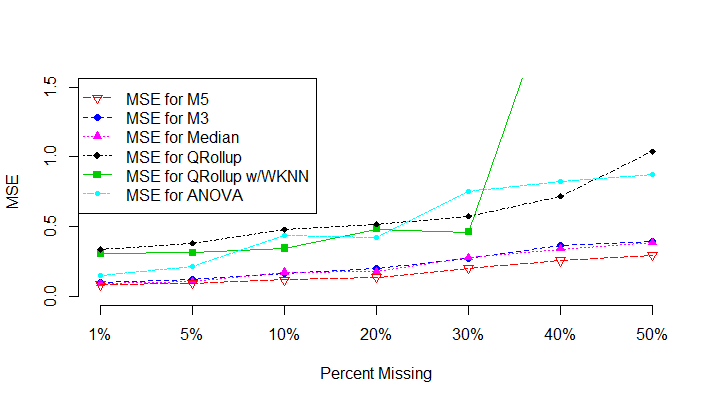}
\end{center}
\caption{MSE computed as the average squared difference between the estimate from a complete case analysis and the estimate from a dataset with simulated intensity dependent missingness.  MSE for QRollup with weighted KNN imputation takes the values 2.34 and 3.31 at 40 and 50 percent missingness respectively.  These values were too extreme to be plotted with the other methods. \label{fig:realdatmse}}
\end{figure}

We explored the effect of intensity dependent missingness on complete case estimates.  The performance demonstrated similar patterns to what we found in the simulation analysis with ratio based methods having far more stable estimates than the average intensity based methods, shown in Figure~\ref{fig:realdatmse}.  

These plots paint a picture consistent with the results from the simulation study.  The M5 model outperforms all other methods.  The method of median ratios and the M3 model have very similar performance.  The ANOVA model and QRollup methods perform comparably to the ratio-based methods until the missingness is increased to around 10\%.  Once missingness hits 40\%, the difference in frameworks becomes substantial, and at 50\% the average MSE from KNNQ is about eleven times higher than that from the M5 model.   

\section{Misspecification of the Missing Data Mechanism}
\begin{figure}
\begin{center}
\includegraphics[width=0.8\textwidth,natwidth=715,natheight=287]{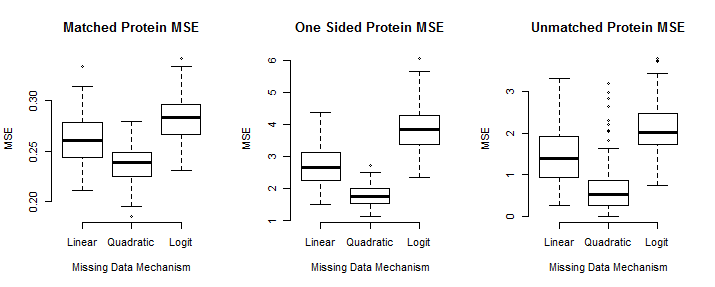}
\end{center}
\caption{MSE computed from 100 simulations utilizing 3 different missing data mechanisms. \label{fig:mechmse}}
\end{figure}

An obvious artificial strength of the simulation study is the use of the same missing data mechanism in both the simulation and the analysis.  The scientific process supports the use of a missing data mechanism in which the probability of a peptide being observed is a monotone increasing function of the intensity. Beyond this basic structure very little evidence exists to suggest the proper shape of this curve.  Our probit model fits the monotonicity requirement however it is not unique in doing so.  To examine the robustness to misspecification of the missing data mechanism we compare estimation results from three different missing data mechanisms; a linear model within a probit function, a quadratic model within a probit function and a linear model within a logit function.  Data was simulated 100 times and for each data set a different set of missing values was simulated according to the three different models.  Simulation parameters were selected so that the overall percentage of missing values would be near 33\%.  As pictured in Figure~\ref{fig:mechmse}, the results suggest that the M5 model is fairly robust to misspecification of the missing data mechanism.  Amongst matched proteins the worst case scenario occurred when the real mechanism was a logit model.  In this case the average MSE increased by 8\% from .26 to .28, which is still 20\% lower than the MSE for the method of medians found in the simulation study.  Misspecification from a quadratic model actually reduced the average MSE by 10\%.  These results seem to suggest that sharper the increase in the probability of observing a peptide the better our model will perform.  For one-sided and unmatched proteins the effects of misspecification were more pronounced.  For one-sided proteins we observed a 42\% increase in average MSE with a probit misspecification and a 34\% decrease for the quadratic model.  For unmatched proteins these changes were 43\% and -53\% respectively.  The increased effect of the missing data mechanism for these proteins should not be surprising since the mechanism plays a larger role in the estimation when no matched pairs are observed.  In the worst case scenario the average MSE for a one-sided Protein from a logit missing data model was 3.87.  With a range of fold changes in the data from roughly -10 to 10 an MSE of 3.87 is highly encouraging as it suggests that even with misspecification the M5 model provides a legitimate way to detect one-sided proteins with large fold changes.    

\section{Conclusion}
We have identified the two fundamental statistical features of mass spectrometry proteomics as matched pairs data and non-ignorable missingness.  Of the two features, ignoring matched pairs appears to be far more detrimental than ignoring the missing data bias.  Not only is the average intensity across peptides difficult to interpret, but the simple method of taking the median ratio greatly outperforms methods based on average intensities in terms of mean squared error.  In turn, relative to the method of medians, our M5 model is capable of improving both the depth and accuracy of the mass spectrometry experiment.  To the best of our knowledge, this is the first model that provides reliable estimation of one-sided proteins and our experiments and sensitivity analysis suggests that these estimates could be valuable even when the missing data mechanism has been misspecified.  A great deal of work can be done to extend this model.  Here we have only demonstrated the importance of peptide level matching with a model that compares two samples from either a SILAC or LFQ experiment.  A different missing data mechanism would be required to fit iTRAQ data and extensions for multiple sample comparisons and sample fractionation are not immediately obvious.  In whatever way these more complicated problems might be solved, the lessons from this study should be incorporated into the solution.  The joint presence of matched pairs data and non-ignorable missingness can greatly inflate the error in estimation procedures that do not explicitly account for both features.

\appendix
\section{APPENDIX: Proofs}
\label{sec:proofs}

\begin{align*}
f_{(Y_{ijk}|\mu_{i},\alpha_{j(i)},Y_{ijk'},\mathbf{\theta},\mathbf{R})}(y_{n})
&=\frac{\prod_{k=n_{1}+1}^{N}{(1-\Phi(a+by_{k}))}\prod_{k=1}^{n_{1}}{\Phi(a+by_{k})}f_{\mathbf{Y_{i}}}
}{\int{\prod_{k=n_{1}+1}^{N}{(1-\Phi(a+by_{k}))}\prod_{k=1}^{n_{1}}{\Phi(a+by_{k})}f_{\mathbf{Y_{i}}}
}dy}\\
&=\frac{(1-\Phi(a+by_{n}))f_{\mathbf{Y_{i}}}
}{\int{(1-\Phi(a+by_{n}))f_{\mathbf{Y_{i}}}}dy}\\
&=\frac{\Phi(-a-by_{n})\exp(-\frac{1}{2\sigma}(y_{n}-\alpha_{n(i)}\pm \frac{\mu_{i}}{2})^2)}
{\int{\Phi(a-by_{n})\exp(-\frac{1}{2\sigma}(y_{n}-\alpha_{n(i)}\pm \frac{\mu_{i}}{2})^2)}dy}\\
&\propto \Phi(-a-by_{n})\exp(-\frac{1}{2\sigma}(y_{n}-\alpha_{n(i)}\pm \frac{\mu_{i}}{2})^2)\\
\end{align*}
Which is the kernel of an extended skew normal distribution defined as

$$
f_{skew}(x)=\frac{\phi(\frac{x-\mu_{x}}{\sigma_{x}})\Phi(\omega\sqrt{1+c^2}+c(\frac{x-\mu_{x}}{\sigma_{x}}))}{\sigma_{x}\Phi(\omega)}
$$
where
$$
\mu_{x}=\alpha_{n(i)}\pm \frac{\mu_{i}}{2},\quad \sigma_{x}=\sqrt{\sigma}
$$
and
$$
\Phi(-a-by_{n})=\Phi(-a-\frac{\sigma_{x}}{\sigma_{x}}b(y_{n}-\mu_{x}+\mu_{x}))=\Phi(-a-\sigma_{x}b(\frac{y_{n}-\mu_{x}}{\sigma_{x}})-b\mu_{x})
$$
Thus,
$$
-b\sigma_{x}=c, \quad \omega\sqrt{1+c^2}=-a-b\mu_{x}
$$
$$
\Rightarrow \omega =\frac{-a-b\mu_{x}}{\sqrt{1+\sigma b^2}}
$$
Therefore,
\begin{align*}
f_{(Y_{ijk}|\mu_{i},\alpha_{j(i)},Y_{ijk'},\mathbf{\theta},\mathbf{R})}(x)&=\frac{\phi(\frac{x-\mu_{x}}{\sqrt{\sigma}})\Phi(\omega\sqrt{1+(-b\sqrt{\sigma})^2}-b\sqrt{\sigma}(\frac{x-\mu_{x}}{\sqrt{\sigma}}))}{\sqrt{\sigma}\Phi(\omega)}\\
&=\frac{\phi(\frac{x-\mu_{x}}{\sqrt{\sigma}})\Phi(-a-bx)}{\sqrt{\sigma}\Phi(\omega)}.
\end{align*}

Similarly 
\begin{align*}
f_{(\mu_{i}|\mathbf{\alpha_{i}},\mathbf{Y},\mathbf{\theta},\mathbf{R})}&\propto f_{\mathbf{Y}|\mu_{i},\mathbf{\alpha}}f_{\mu_{i}}\\
&\propto \exp(-\frac{1}{2\tau}(\mu_{i}-\beta_{\mu})^2) \prod_{j=1}^{m_{i}}{\exp\left(-\frac{1}{2\sigma}\left((y_{ij1}-\alpha_{j(i)}-\frac{\mu_{i}}{2})^2+(y_{ij2}-\alpha_{j(i)}+\frac{\mu_{i}}{2})^2 \right)\right)}\\
&=\exp\left(-\frac{1}{2\tau}(\mu_{i}-\beta_{\mu})^2 + \sum_{j=1}^{m_{i}}{-\frac{1}{2\sigma}\left((y_{ij1}-\alpha_{j(i)}-\frac{\mu_{i}}{2})^2+(y_{ij2}-\alpha_{j(i)}+\frac{\mu_{i}}{2})^2 \right)}\right)\\
&=\exp\left(-\frac{1}{2}\left(\frac{1}{\tau}\mu_{i}^2 - \frac{2\beta_{\mu}}{\tau}\mu_{i}+\frac{1}{\sigma} \sum_{j=1}^{m_{i}} \frac{1}{2}\mu_{i}^{2} -\mu_{i}(y_{ij1}-\alpha_{j(i)})+\mu_{i}(y_{ij2}-\alpha_{j(i)})  \right) + C\right)\\
\end{align*}
for some constant C.
\begin{align*}
&\propto \exp\left(-\frac{1}{2}\left(\mu_{i}^2(\frac{1}{\tau}+\frac{m_{i}}{2\sigma})-\mu_{i}\left(\frac{2\beta_{\mu}}{\tau} +\frac{1}{\sigma}\sum{(y_{ij1}-y_{ij2})} \right) \right)  \right)\\
&=\exp\left(-\frac{1}{2(\frac{2\sigma\tau}{2\sigma+m_{i}\tau})}\left(\mu_{i}^2-\mu_{i}\frac{2\sigma\tau}{2\sigma+m_{i}\tau}\left(\frac{2\beta_{\mu}}{\tau}+\frac{1}{\sigma}\sum{(y_{ij1}-y_{ij2})}\right) \right) \right)\\
&=\exp\left(-\frac{1}{2(\frac{2\sigma\tau}{2\sigma+m_{i}\tau})}\left(\mu_{i}^2-\mu_{i}(\frac{2\beta_{\mu}\sigma+\tau\sum{(y_{ij1}-y_{ij2})}}{\sigma+\frac{m_{i}\tau}{2}}) \right) \right)\\
&\propto\exp\left(-\frac{1}{2(\frac{2\sigma\tau}{2\sigma+m_{i}\tau})}\left(\mu_{i}-(\frac{\beta_{\mu}\sigma+\frac{\tau}{2}\sum{(y_{ij1}-y_{ij2})}}{\sigma+\frac{m_{i}\tau}{2}}) \right)^2 \right)\\
\end{align*}
Therefore,
$$
(\mu_{i}|\mathbf{\alpha_{i}},\mathbf{Y},\mathbf{\theta},\mathbf{R})\sim N\left(\frac{\beta_{\mu}\sigma+\frac{\tau}{2}\sum_{j}{(y_{ij1}-y_{ij2})}}{\sigma+\frac{m_{i}\tau}{2}},\frac{\sigma\tau}{\sigma+\frac{m_{i}\tau}{2}}\right).
$$

The distribution for $\alpha$ given everything else can be derived in a similar manner.

\begin{align*}
f_{(\alpha_{j(i)}|\mu_{i},\mathbf{Y},\mathbf{\theta},\mathbf{R})}&\propto f_{\mathbf{Y}|\mu_{i},\mathbf{\alpha}}f_{\alpha_{j(i)}}\\
&=\exp(-\frac{1}{2\xi}(\mu_{i}-\beta_{\alpha})^2) \exp\left(-\frac{1}{2\sigma}\left((y_{i,j,1}-\alpha_{j(i)}-\frac{\mu_{i}}{2})^2+(y_{ij2}-\alpha_{j(i)}+\frac{\mu_{i}}{2})^2 \right)\right)\\
&=\exp(-\frac{1}{2}(\frac{1}{\xi}(\alpha_{j(i)}^2-2\beta_{\alpha}\alpha_{j(i)}) +\frac{1}{\sigma}\left(2\alpha_{j(i)}^2-2\alpha_{j(i)}(y_{ij1}-\frac{\mu_{i}}{2})-2\alpha_{j(i)}(y_{ij2}+\frac{\mu_{i}}{2}) \right))+C)\\
\end{align*}
for some constant C
\begin{align*}
&\propto\exp(-\frac{1}{2}(\frac{1}{\xi}(\alpha_{j(i)}^2-2\beta_{\alpha}\alpha_{j(i)}) +\frac{2}{\sigma}\left(\alpha_{j(i)}^2-\alpha_{j(i)}(y_{ij1}+y_{ij2}) \right)))\\
&=\exp(-\frac{1}{2}(\alpha_{j(i)}^2(\frac{1}{\xi}+\frac{2}{\sigma})-\alpha_{j(i)}(\frac{2\beta_{\alpha}}{\xi}+2\frac{y_{ij1}+y_{ij2}}{\sigma})))\\
&=\exp(-\frac{1}{2\frac{\xi\sigma}{\sigma+2\xi}}(\alpha_{j(i)}^2-\alpha_{j(i)}\frac{\xi\sigma}{\sigma+2\xi}(\frac{2\beta_{\alpha}}{\xi}+2\frac{y_{ij1}+y_{ij2}}{\sigma})))\\
&=\exp(-\frac{1}{2\frac{\xi\sigma}{\sigma+2\xi}}(\alpha_{j(i)}^2-\alpha_{j(i)}(\frac{2\beta_{\alpha}\sigma+2\xi(y_{ij1}+y_{ij2})}{\sigma+2\xi})))\\
&=\exp(-\frac{1}{2\frac{\xi\sigma}{\sigma+2\xi}}(\alpha_{j(i)}-(\frac{\beta_{\alpha}\sigma+\xi(y_{ij1}+y_{ij2})}{\sigma+2\xi}))^2)\\
\end{align*}
Therefore,
$$
(\alpha_{j(i)}|\mu_{i},\mathbf{Y},\mathbf{\theta},\mathbf{R})\sim N\left(\frac{\beta_{\alpha}\sigma+\xi(y_{ij1}+y_{ij2})}{\sigma+2\xi},\frac{\xi\sigma}{\sigma+2\xi} \right)
$$

\bigskip
\begin{center}
{\large\bf SUPPLEMENTARY MATERIAL}
\end{center}

\begin{description}

\item[R-code for implementing the M5 model] R code for performing M5 and the other methods described in the article.

\item[Tumor data set:] Data set used in the sensitivity analysis (.csv file)

\end{description}

\bibliographystyle{Chicago}

\bibliography{M5Revised}

\begin{thebibliography}{}

\bibitem[\protect\citeauthoryear{Azzalini and Capitanio}{Azzalini and
  Capitanio}{2014}]{Azzalini2014}
Azzalini, A. and A.~Capitanio (2014).
\newblock {\em {The skew-normal and related families}}.
\newblock Cambridge: Cambridge University Press.

\bibitem[\protect\citeauthoryear{Beardsley, Kowbel, Lataxes, Mannino, Xin, Kim,
  Collins, and Brown}{Beardsley et~al.}{2003}]{Beardsley2003}
Beardsley, D.~I., D.~Kowbel, T.~A. Lataxes, J.~M. Mannino, H.~Xin, W.-J. Kim,
  C.~Collins, and K.~D. Brown (2003, November).
\newblock {Characterization of the novel amplified in breast cancer-1 (NABC1)
  gene product.}
\newblock {\em Experimental cell research\/}~{\em 290\/}(2), 402--13.

\bibitem[\protect\citeauthoryear{Cox, Hein, Luber, Paron, Nagaraj, and
  Mann}{Cox et~al.}{2014}]{Cox2014}
Cox, J., M.~Y. Hein, C.~A. Luber, I.~Paron, N.~Nagaraj, and M.~Mann (2014,
  June).
\newblock {MaxLFQ allows accurate proteome-wide label-free quantification by
  delayed normalization and maximal peptide ratio extraction.}
\newblock {\em Molecular \& cellular proteomics : MCP\/}~{\em 13\/}(9),
  2513--2526.

\bibitem[\protect\citeauthoryear{Cox and Mann}{Cox and Mann}{2008}]{Cox2008}
Cox, J. and M.~Mann (2008).
\newblock {MaxQuant enables high peptide identification rates, individualized
  p.p.b.-range mass accuracies and proteome-wide protein quantification.}
\newblock {\em Nature biotechnology\/}~{\em 26\/}(12), 1367--1372.

\bibitem[\protect\citeauthoryear{Dabney and Storey}{Dabney and
  Storey}{2007}]{Dabney2007}
Dabney, A.~R. and J.~D. Storey (2007).
\newblock {A new approach to intensity-dependent normalization of two-channel
  microarrays}.
\newblock {\em Biostatistics\/}~{\em 8}, 128--139.

\bibitem[\protect\citeauthoryear{de~Brevern, Hazout, and Malpertuy}{de~Brevern
  et~al.}{2004}]{DeBrevern2004}
de~Brevern, A.~G., S.~Hazout, and A.~Malpertuy (2004).
\newblock {Influence of microarrays experiments missing values on the stability
  of gene groups by hierarchical clustering.}
\newblock {\em BMC bioinformatics\/}~{\em 5}, 114.

\bibitem[\protect\citeauthoryear{Eidhammer, Flikka, Martens, and
  Mikalsen}{Eidhammer et~al.}{2008}]{Eidhammer2008}
Eidhammer, I., K.~Flikka, L.~Martens, and S.-O. Mikalsen (2008).
\newblock {\em {Computational Methods for Mass Spectrometry Proteomics}}.
\newblock John Wiley \& Sons.

\bibitem[\protect\citeauthoryear{Gelman, Carlin, Stern, and Rubin}{Gelman
  et~al.}{2004}]{Gelman2004}
Gelman, A., J.~B. Carlin, H.~S. Stern, and D.~B. Rubin (2004).
\newblock {\em {Bayesian Data Analysis Second Edition}}, Volume~1.
\newblock Chapman \& Hall/CRC.

\bibitem[\protect\citeauthoryear{Karpievitch, Stanley, Taverner, Huang, Adkins,
  Ansong, Heffron, Metz, Qian, Yoon, Smith, and Dabney}{Karpievitch
  et~al.}{2009}]{Karpievitch2009}
Karpievitch, Y., J.~Stanley, T.~Taverner, J.~Huang, J.~N. Adkins, C.~Ansong,
  F.~Heffron, T.~O. Metz, W.~J. Qian, H.~Yoon, R.~D. Smith, and A.~R. Dabney
  (2009).
\newblock {A statistical framework for protein quantitation in bottom-up
  MS-based proteomics}.
\newblock {\em Bioinformatics\/}~{\em 25\/}(16), 2028--2034.

\bibitem[\protect\citeauthoryear{Koopmans, Cornelisse, Heskes, and
  Dijkstra}{Koopmans et~al.}{2014}]{Koopmans2014}
Koopmans, F., L.~N. Cornelisse, T.~Heskes, and T.~M.~H. Dijkstra (2014,
  September).
\newblock {Empirical bayesian random censoring threshold model improves
  detection of differentially abundant proteins.}
\newblock {\em Journal of proteome research\/}~{\em 13\/}(9), 3871--80.

\bibitem[\protect\citeauthoryear{Lesur and Domon}{Lesur and
  Domon}{2015}]{Lesur2015}
Lesur, A. and B.~Domon (2015, March).
\newblock {Advances in high-resolution accurate mass spectrometry application
  to targeted proteomics.}
\newblock {\em Proteomics\/}~{\em 15\/}(5-6), 880--90.

\bibitem[\protect\citeauthoryear{Lucas, Thompson, Dubois, McCarthy, Tillmann,
  Thompson, Shire, Hendrickson, Dieguez, Goldman, Schwarz, Patel, McHutchison,
  and Moseley}{Lucas et~al.}{2012}]{Lucas2012}
Lucas, J.~E., J.~Thompson, L.~G. Dubois, J.~McCarthy, H.~Tillmann, A.~Thompson,
  N.~Shire, R.~Hendrickson, F.~Dieguez, P.~Goldman, K.~Schwarz, K.~Patel,
  J.~McHutchison, and M.~Moseley (2012).
\newblock {Metaprotein expression modeling for label-free quantitative
  proteomics}.
\newblock {\em BMC Bioinformatics\/}~{\em 13\/}(1), 74.

\bibitem[\protect\citeauthoryear{Luo, Colangelo, Sessa, and Zhao}{Luo
  et~al.}{2009}]{Luo2009}
Luo, R., C.~M. Colangelo, W.~C. Sessa, and H.~Zhao (2009, November).
\newblock {Bayesian Analysis of iTRAQ Data with Nonrandom Missingness:
  Identification of Differentially Expressed Proteins.}
\newblock {\em Statistics in biosciences\/}~{\em 1\/}(2), 228--245.

\bibitem[\protect\citeauthoryear{Oberg and Mahoney}{Oberg and
  Mahoney}{2012}]{Oberg2012}
Oberg, A.~L. and D.~W. Mahoney (2012).
\newblock {Statistical methods for quantitative mass spectrometry proteomic
  experiments with labeling.}
\newblock {\em BMC bioinformatics\/}~{\em 13 Suppl 1\/}(Suppl 16), S7.

\bibitem[\protect\citeauthoryear{Polpitiya, Qian, Jaitly, Petyuk, Adkins, Camp,
  Anderson, and Smith}{Polpitiya et~al.}{2008}]{Polpitiya2008}
Polpitiya, A.~D., W.-J. Qian, N.~Jaitly, V.~A. Petyuk, J.~N. Adkins, D.~G.
  Camp, G.~A. Anderson, and R.~D. Smith (2008, July).
\newblock {DAnTE: a statistical tool for quantitative analysis of -omics data.}
\newblock {\em Bioinformatics (Oxford, England)\/}~{\em 24\/}(13), 1556--8.

\bibitem[\protect\citeauthoryear{Ross, Huang, Marchese, Williamson, Parker,
  Hattan, Khainovski, Pillai, Dey, Daniels, Purkayastha, Juhasz, Martin,
  Bartlet-Jones, He, Jacobson, and Pappin}{Ross et~al.}{2004}]{Ross2004}
Ross, P.~L., Y.~N. Huang, J.~N. Marchese, B.~Williamson, K.~Parker, S.~Hattan,
  N.~Khainovski, S.~Pillai, S.~Dey, S.~Daniels, S.~Purkayastha, P.~Juhasz,
  S.~Martin, M.~Bartlet-Jones, F.~He, A.~Jacobson, and D.~J. Pappin (2004,
  December).
\newblock {Multiplexed protein quantitation in Saccharomyces cerevisiae using
  amine-reactive isobaric tagging reagents.}
\newblock {\em Molecular \& cellular proteomics : MCP\/}~{\em 3\/}(12),
  1154--69.

\bibitem[\protect\citeauthoryear{Sandin, Krogh, Hansson, and Levander}{Sandin
  et~al.}{2011}]{Sandin2011}
Sandin, M., M.~Krogh, K.~Hansson, and F.~Levander (2011, March).
\newblock {Generic workflow for quality assessment of quantitative label-free
  LC-MS analysis.}
\newblock {\em Proteomics\/}~{\em 11\/}(6), 1114--24.

\bibitem[\protect\citeauthoryear{Schliekelman and Liu}{Schliekelman and
  Liu}{2014}]{Schliekelman2014}
Schliekelman, P. and S.~Liu (2014, February).
\newblock {Quantifying the effect of competition for detection between
  coeluting peptides on detection probabilities in mass-spectrometry-based
  proteomics.}
\newblock {\em Journal of proteome research\/}~{\em 13\/}(2), 348--61.

\bibitem[\protect\citeauthoryear{Taylor, Leiserowitz, and Kim}{Taylor
  et~al.}{2013}]{Taylor2013}
Taylor, S.~L., G.~S. Leiserowitz, and K.~Kim (2013, December).
\newblock {Accounting for undetected compounds in statistical analyses of mass
  spectrometry 'omic studies.}
\newblock {\em Statistical applications in genetics and molecular
  biology\/}~{\em 12\/}(6), 703--22.

\bibitem[\protect\citeauthoryear{Tekwe, Carroll, and Dabney}{Tekwe
  et~al.}{2012}]{Tekwe2012}
Tekwe, C.~D., R.~J. Carroll, and A.~R. Dabney (2012).
\newblock {Application of survival analysis methodology to the quantitative
  analysis of LC-MS proteomics data}.
\newblock {\em Bioinformatics\/}~{\em 28\/}(15), 1998--2003.

\bibitem[\protect\citeauthoryear{Troyanskaya, Cantor, Sherlock, Brown, Hastie,
  Tibshirani, Botstein, and Altman}{Troyanskaya et~al.}{2001}]{Troyanskaya2001}
Troyanskaya, O., M.~Cantor, G.~Sherlock, P.~Brown, T.~Hastie, R.~Tibshirani,
  D.~Botstein, and R.~B. Altman (2001, June).
\newblock {Missing value estimation methods for DNA microarrays}.
\newblock {\em Bioinformatics\/}~{\em 17\/}(6), 520--525.

\end{thebibliography}
\end{document}